# Thermodynamic evidence for the Fulde-Ferrell-Larkin-Ovchinnikov state in the $KFe_2As_2$ superconductor

Chang-woo Cho[1], Jonathan Haiwei Yang[1], Junying Shen[1], Thomas Wolf[2] and Rolf Lortz[1χ]

[1]*Department of Physics, The Hong Kong University of Science and Technology, Clear Water Bay, Kowloon, Hong Kong*

[2]*Institute for Solid State Physics, Karlsruhe Institute of Technology, PO Box 3640, 76021 Karlsruhe, Germany*

We have investigated the magnetic phase diagram near the upper critical field of $KFe_2As_2$ by magnetic torque and specific heat experiments, using a high-resolution piezo-rotary positioner to precisely control the parallel orientation of the magnetic field with respect to the FeAs layers. We observe a clear double transition when the field is oriented strictly in-plane, and a characteristic upturn of the upper critical field line well beyond the Pauli limit at 4.7 T. This provides firm evidence that an FFLO state is realized in this iron-based $KFe_2As_2$ superconductor.

While the upper critical field ($H_{c2}$) of type-II superconductors is usually determined by the orbital limit for superconductivity [1], there are rare cases in which the Pauli paramagnetic limit, at which the Zeeman split Fermi surfaces suppress Cooper pairing, occurs instead [2,3]. When a superconductor approaches the latter, an exotic superconducting state was predicted by Fulde, Ferrell, Larkin and Ovchinnikov [4,5]. Under certain conditions, a superconductor could overcome the Pauli limit by forming the 'FFLO' state in which the Cooper pairs have a finite center-of-mass momentum. Cooper pairing between the Zeeman split Fermi surfaces is possible only with an oscillating component of the order parameter amplitude in real space. Only a few observations of the FFLO state have been reported to date despite its high technological relevance for high magnetic field applications. The Q-phase in the heavy fermion superconductor $CeCoIn_5$ is regarded as a possible realization [6-8], although the superconductivity coexists with an incommensurate spin-density wave [8]. In some layered organic superconductors, evidence has been provided for an FFLO state without magnetic order [9-16]. The rarity of the FFLO state is due to the strict requirements on its occurrence. It needs a very large Ginzburg-Landau parameter $\kappa \equiv \lambda/\xi \gg 1$ ($\lambda$: penetration depth, $\xi$: coherence length), and a Maki parameter $\alpha$ greater than 1.85 [17,18]. Only in this case does the Pauli limit occur below the orbital limit. In addition, the superconductor must be in the clean limit [19,20]. Low dimensionality and anisotropic Fermi surfaces can further stabilize the FFLO state [21]. The $H_{c2}$ line in the magnetic ($H$-$T$) phase diagram shows a characteristic saturation at the Pauli limit, where the second-order transition (SOT) changes into a discontinuous first-order transition (FOT). However, as soon an FFLO state is formed, a characteristic increase of $H_{c2}$ occurs to fields beyond the Pauli limi.

$KFe_2As_2$ is the overdoped end member of the $Ba_{1-x}K_xFe_2As_2$ family of the multiband Fe-based superconductors with $T_c$ of ~3.4 K. In recent magnetostriction experiments it was found that $H_{c2}$ is Pauli limited for fields applied parallel to the FeAs layers [22,23]. However,

χ Corresponding author: lortz@ust.hk

no sign of an FFLO state has so far been reported. One possible reason may be that its observation requires precision in the parallel field orientation better than 1° [11]. Any finite perpendicular field component may generate the formation of a vortex state and suppresses the FFLO state. $KFe_2As_2$ is a superconductor in the very clean limit with $l/\xi{\sim}15$ and a large Maki parameter of 1.7~3.4 [22,23] with the possibility of a nodal *d*-wave order parameter [24]. In this letter we provide thermodynamic evidence for an FFLO state in $KFe_2As_2$ by investigating its *H*-*T* phase diagram close to the Pauli limit with a perfect in-plane field orientation by means of magnetic torque ($\tau$) and specific heat ($C_p$) experiments. Only for accurate alignment we observe a characteristic upturn of the $H_{c2}$ line beyond the Pauli limit, as well as the double transition from the homogeneous superconducting state via the FFLO state to the normal state.

The experiments were performed on a $KFe_2As_2$ single crystal. Details of sample growth and characterization can be found elsewhere [22,23]. The thin foil-like sample was flattened carefully between two glass plates. All experiments were carried out on an Attocube ANR51 piezo rotary stepper positioner, which offers a millidegree fine positioning resolution. The rotator was mounted on a $^3$He probe in a 15 T magnet cryostat. The torque was measured using a home-made capacitance torque magnetometer with a General Radio 1615A capacitance bridge in combination with a SR830 digital lock-in amplifier. The specific heat was measured with a homemade AC calorimeter [25,26] using an SR830 digital lock-in amplifier. Since the transition into the FFLO state often occurs as a nearly horizontal line in the *H*-*T* phase diagram, we have carried out all experiments at a fixed base temperature as a function of field, while the temperature was periodically modulated with a milli-Kelvin amplitude. The calorimeter chip is suspended on a pair of thin nylon wires that provide firm support at low temperature to avoid torque artefacts.

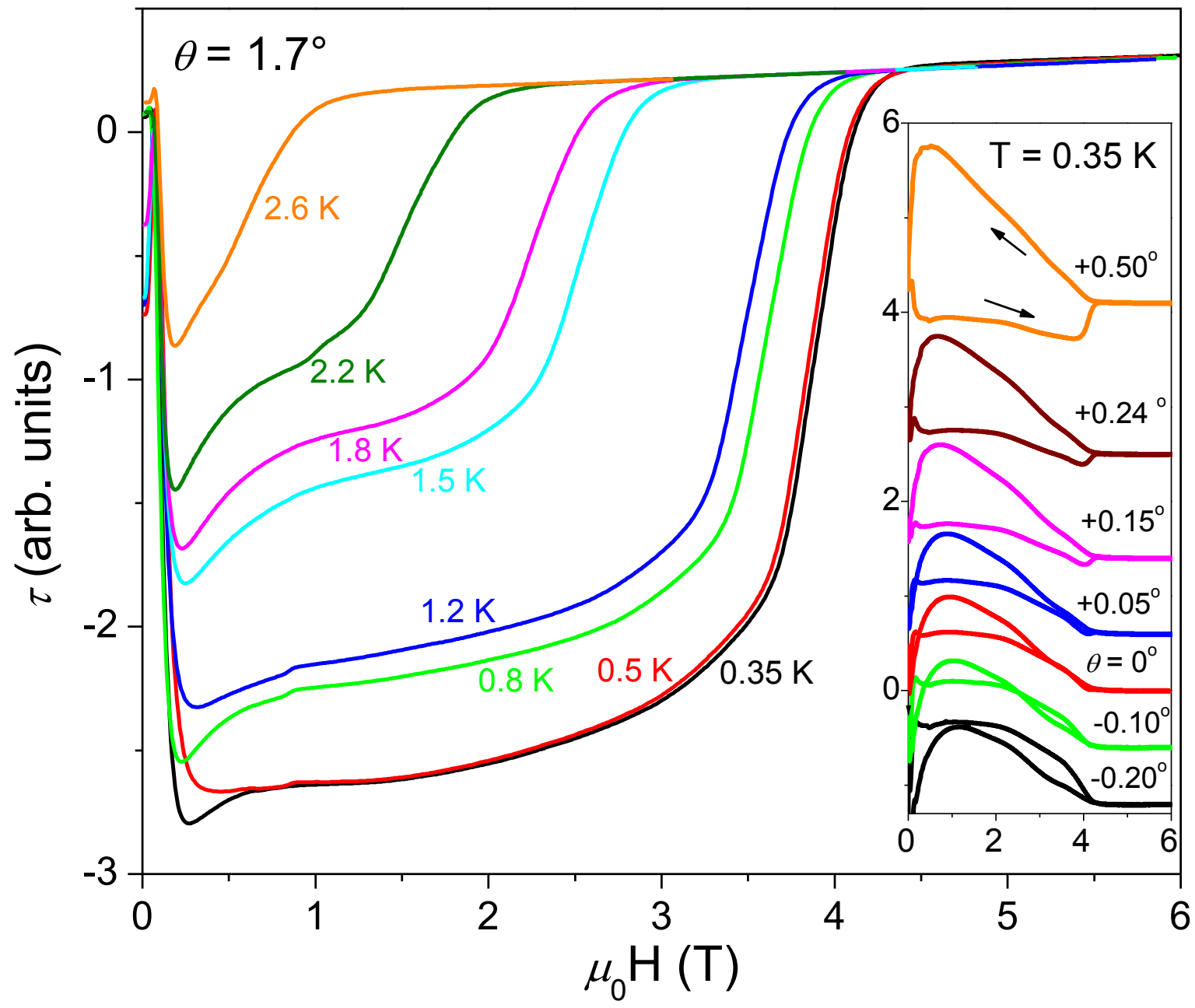

FIG. 1. The magnetic torque as a function of the applied magnetic field of $KFe_2As_2$, measured at fixed temperatures. The field was aligned with a small angle of $\theta = 1.7°$ to the FeAs layers. The data were measured with increasing field. Inset: Magnetic torque at various angles close to $\theta = 0°$. The linear normal state background was removed here for clarity.

In Fig. 1 we show $\tau$ data measured at various fixed temperatures as a function of field with a small $\theta = 1.7°$ misalignment of the field with respect to the FeAs layers. The torque shows the characteristic change of a kink-like SOT at $H_{c2}$ (2.6 K) to a jump-like FOT at lower temperatures. $H_{c2}$ between 0.5 K and 0.35 K remains almost unchanged. These are clear signs that the Pauli limit is approached, and the behavior is similar to magnetostriction data reported previously [23]. Note that the jump in the torque is mostly arising from the abrupt decay of irreversible screening currents when approaching the Pauli limit, but may occur at slightly lower fields than $H_{c2}$ in a purely thermodynamic quantity. $H_{c2}$ should be rather determined by the first deviation from the normal state background, which may be represented by a smaller reversible component. No indication for an FFLO state is observed here and the superconductivity vanishes in form of this FOT.

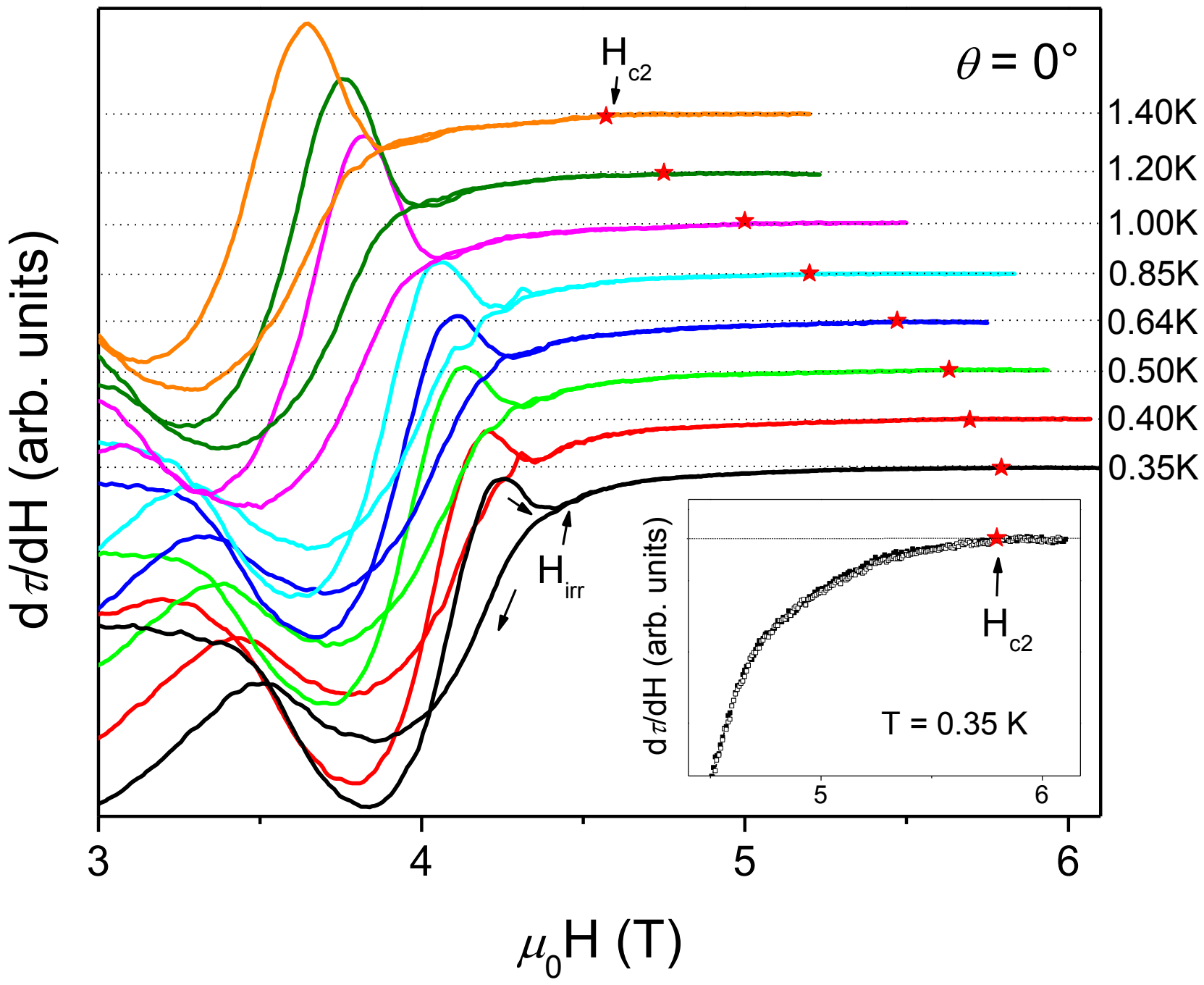

FIG. 2. The first-order derivative of the magnetic torque as a function of the applied field at different temperatures for the field strictly parallel to the FeAs layers ($\theta = 0°$). The stars mark the upper onset of the superconducting signal and the dotted lines the constant normal state value. Vertical offsets have been added to the data. The inset shows an enlarged region of the $T = 0.35$ K data at $H_{c2}$.

In the next step we aligned the FeAs layers precisely parallel to the field by minimizing the torque at 0.35 K. A sequence of measurements is shown in the inset of Fig. 1. It can be seen that $\tau$ becomes smallest at $\theta = 0°$, and also the hysteresis almost disappears. The normal state background is completely linear, but a small deviation up to 5.8 T is seen at $\theta = 0°$, as we will demonstrate using the first field derivative of the torque at $\theta = 0°$ in Fig. 2. We define $H_{c2}$ from a deviation point from the constant normal state value (dotted lines), as illustrated in the inset. The main change in $\tau$ initiates at a characteristic irreversibility field ($H_{irr}$) below which the two branches taken on increasing and decreasing fields begin to deviate. A reversible superconducting contribution in the form of a gradual downturn is observed far above $H_{irr}$, which extends up to 5.8 T at 0.35 K. Note that the torque is directly related to the anisotropic component of the magnetization, which in the absence of irreversibility is a bulk thermodynamic method. The onset corresponds to the start of Cooper pairing and thus best represents $H_{c2}$. Additional data showing $\tau$ measured at 3 different angles can be found in the supplementary materials. $H_{c2}$ is increased from 5 T to 5.8 T when the angle varies from 1.7° to 0°. This may indicate that an FFLO state associated with this reversible tail in torque goes far beyond the Pauli limit [23], but develops only for accurate parallel field orientation, while at an angle of more than 1° it disappears and the superconductor remains Pauli limited.

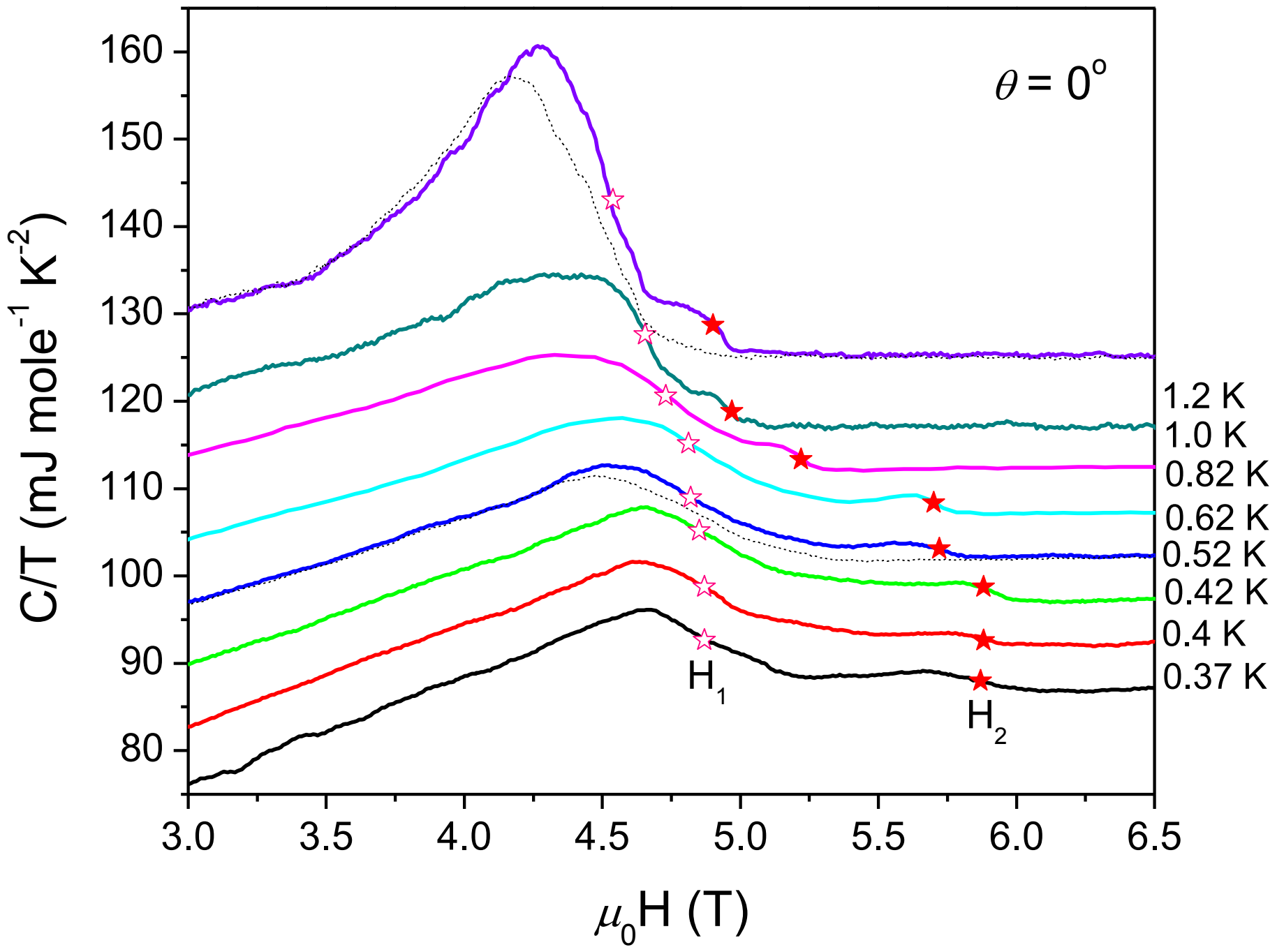

FIG. 3. Specific heat $C_p/T$ measured as a function of magnetic field aligned with $\theta = 0°$ with respect to the FeAs layers at different fixed temperatures. Filled stars mark the additional phase transitions at $H_2$ above the main transitions at $H_1$ (empty stars). Vertical offsets were added for clarity. The additional dashed lines were measured under identical conditions at 0.52 K and 1.2 K but with a misalignment angle $\theta = 3°$ instead.

In order to further test the possible existence of an FFLO state, we shall now investigate the specific heat in the vicinity of $H_{c2}$. Fig. 3a shows the field dependence of $C_p/T$ at $\theta = 0°$ at various fixed temperatures. The 1.2 K data show a main transition at characteristic fields $H_1$ in the form of a broad triangular peak at 4.3 T. At lower temperatures, this transition is transformed into a broad step-like anomaly with midpoint which increases to 4.7 T at 0.37 K. In all data, another small step-like phase transition anomaly occurs above the main transition and reaches a field $H_2$ (= $H_{c2}$) = 5.8 T at 0.37 K. The characteristic fields $H_2$ at the various temperatures agree well with the onset of the superconducting $\tau$ contribution. We added two sets of data at 0.52 K and 1.2 K in which we had a small misalignment angle $\theta = 3°$ for comparison. The 3° data show only one transition and the additional step at $H_2$ is completely missing. In addition, the main transition is slightly shifted to lower fields. A pronounced double transition only develops for a precisely parallel field alignment. The transition at $H_1$ corresponds to $H_{irr}$ in $\tau$ and initiates the rapid increase in the torque magnitude at lower fields.

Note that it is difficult to judge the phase transition order without careful consideration. $C_p$ data as a function of the field are rarely shown and should not be confused with the magnetocaloric effect [26,27]. The latter represents the calorimetric response with respect of a field change, while $C_p$ is measured in response to a temperature modulation during a field sweep. Both transitions at $H_1$ and at $H_2$ appear, at first sight, as a SOT in the form of a characteristic jump. However, the transition at $H_1$ sharpens at 1.2 K and evolves into a more symmetrical form, which is probably the signature of a FOT. The absence of latent heat could be a result of the low temperature and the vanishing slope of the phase boundaries, which make the entropy differences between the phases disappear, or due to a limitation of the AC technique which is known to underestimate latent heat [28].

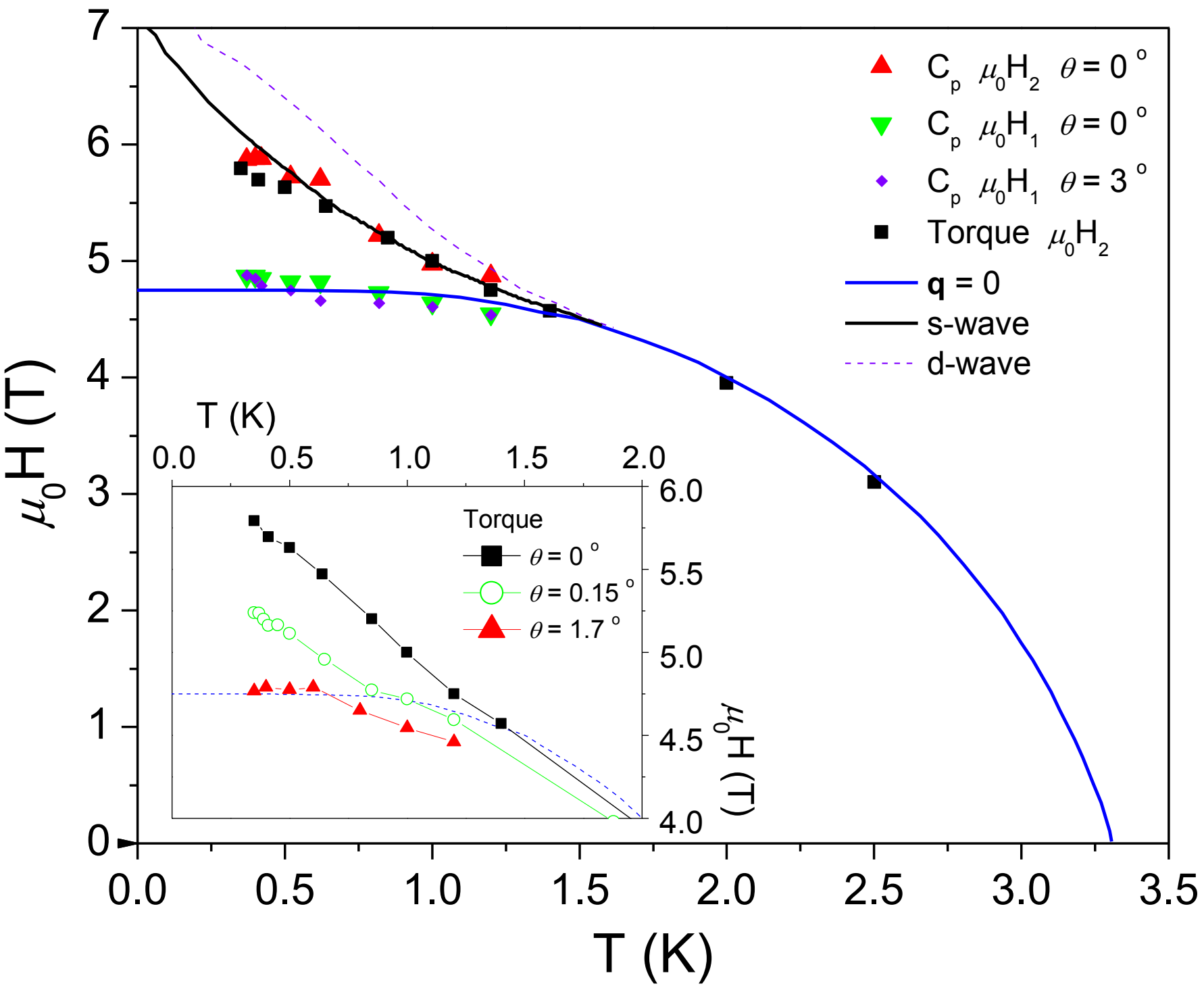

FIG. 4. *H*-*T* phase diagram of $KFe_2As_2$ from magnetic torque and specific heat. $\mu_0H_1$ represents the transition between the homogeneous superconducting state (**q** = 0) and the FFLO state. The upper transition at $\mu_0H_2$ separates the FFLO state from the normal state. The additional lines represents the $H_{c2}(T)$ lines predicted for a purely Pauli-limited superconductor (**q** = 0) [29], and for an in-plane isotropic s-wave and a d-wave superconductor with FFLO state [30]. The inset shows an enlargement with the upper onset of the superconducting contribution in torque for different field orientations (0°, 0.15° and 1.7°).

In Fig. 4, we compile an *H*–*T* phase diagram from our $\tau$ and $C_p$ data. There are numerous theoretical descriptions of the $H_{c2}$ transition with orbital and Zeeman pair breaking effects in clean superconductors [29-32]. Gurevich provided a recent model of the $H_{c2}$ transition in anisotropic multiband superconductors with and without the FFLO state [29], and we include his $H_{c2}$ prediction for the purely Pauli-limited case in Fig. 4. It agrees well with our data in which we had a small field misalignment. Meanwhile, for the perfect parallel field orientation, both $\tau$ and $C_p$ first follow the line of the Pauli-limited case at temperatures above ~1.5 K, but then the $H_2$ line gradually turns up to higher fields. In addition, the characteristic double transition in $C_p$ is observed as a signature of the transition sequence from the ordinary superconducting phase, via the FFLO state ($H_1$), to the normal conducting state ($H_2$). The main $C_p$ transition at $H_1$ remains at the Pauli limit, which therefore initiates the field-driven transition into the FFLO state. The phase diagram thus shows all the characteristic features expected for the formation of the FFLO state, with the FFLO boundary being represented by the lines marked as $H_1$ and $H_2$ as the lower and upper boundary, respectively.

The $H_{c2}$ transition with the FFLO state for mixtures of s- and d-wave superconductors has been predicted in Ref 30. An in the plane isotropic s-wave superconductor should show a pronounced convex upturn at low temperatures, and in fact this model fits best our data. The *H*-*T* phase diagram thus does not reflect the unconventional multiband structure of Fe-based superconductors. This may be due to the fact that in $KFe_2As_2$ the electronic band structure contains only hole bands in form of $\alpha$, $\beta$, $\zeta$ and $\varepsilon$ bands [33]. However, at the lowest temperatures our data deviates slightly with a saturating concave trend. The d-wave model describes such change of curvature well, but the FFLO phase is too large. A mixture of s and d-wave would further enlarge it. The saturation behavior was also predicted for anisotropic single-band and multiband s-wave order parameters [29], but the predicted FFLO phases are too small. However, a combination of multiband superconductivity with d-wave order parameter, could cancel these effects and mimic a simple s-wave behavior. The effect of multiband superconductivity with finite intraband pairing interaction on an FFLO state in $KFe_2As_2$ was predicted to cause a separation of the FFLO phase into a $Q_1$ and a $Q_2$ phase with different order parameter modulation vectors [31]. The $Q_2$ phase is expected to occur at lower temperatures and is separated by a first-order transition from the $Q_1$ phase at higher fields and temperatures. While we do not see evidence of an additional transition, it is possible that it may occur at lower temperatures. The *H*-*T* phase diagram of $KFe_2As_2$ is very similar to κ-(BEDT-TTF)$_2$Cu(NCS)$_2$, although the experimental signatures are quite different. In κ-(BEDT-TTF)$_2$Cu(NCS)$_2$, there is only a small $\tau$ [11] and $C_p$ [27] anomaly at $H_1$, while the greatest change between the superconducting and the normal state occurs at $H_2$. This is different in $KFe_2As_2$, for which the main transition anomaly in $C_p$ occurs at $H_1$, with only a small superconducting contribution of ~15% of the total $C_p$ jump persisting up to $H_2$. This could indicate that the FFLO state forms only in one band [34]. In fact, it was reported that the gap in the $\varepsilon$ band is much greater than the others [35]. Therefore, while the smaller gaps are largely suppressed in high fields, the $\varepsilon$ band would be the only one to remain gapped up to $H_p$ and thus dictates the characteristics of the FFLO state.

The transition into the FFLO state occurs at 4.7 T, while the theoretical weak-coupling paramagnetic limit $H_p(0) = 1.84T_c \approx 6$ T is slightly higher. A more precise value is obtained from $H_P(0) = \Delta_0/(\sqrt{2}\mu_B) \sim 7.4$ T, wherein an average value $\Delta_0/k_BT_c \approx 1.9$ is used [35]. The latter is contributed by the $\varepsilon$ gap, which dominates the zero-field $C_p$ jump at $T_c$. However, these formulae likely overestimate the Pauli limit because of the large normal-state Pauli susceptibility of $KFe_2As_2$ [36]. Using $H_P(0) = H_c(0)/(\sqrt{\chi_n - \chi_s}) \sim 7.4$ [37] ($\chi_n$ and $\chi_s$ are the normal-state and superconducting spin susceptibilities, respectively), yields a value of $H_p(0) = 3.6$ T [22], which may be somewhat increased by the multiband superconductivity. A more exotic possibility is that, similar to $CeCoIn_5$ [8], a competing magnetic order coexists in $KFe_2As_2$ with the FFLO state thus pushing the paramagnetic limit to lower fields. This may be linked to its strongly correlated nature [33] with highly renormalized values of the Sommerfeld coefficient and $\chi_n$ [37]. Note that even with $Ba_{0.07}K_{0.93}Fe_2As_2$, a FOT with a certain upturn of the $H_{c2}$ line was observed, although the latter was interpreted differently [38]. In addition, evidence for an unusual field-induced superconducting phase was reported for FeSe [39]. Therefore, it would be interesting to perform similar experiments on other Fe-based superconductors.

R. L. would like to thank J. Wosnitza, K. Grube, T. Terashima and A. Ptok for their valuable suggestions.

[1] L. P. Gor'kov, The Critical Supercooling Field in Superconductivity Theory, Sov. Phys. JETP 10, 593 (1960).

[2] A. K. Clogston, Upper Limit for the Critical Field in Hard Superconductors, Phys. Rev. Lett. **9**, 266 (1962).

[3] B. S. Chandrasekhar, A note on the maximum critical field of high-field superconductors, Appl. Phys. Lett. **1**, 7 (1962).

[4] P. Fulde, R. A. Ferrell, Superconductivity in a Strong Spin-Exchange Field, Phys. Rev. **135**, A550 (1964).

[5] A. I. Larkin, Yu. N. Ovchinnikov, Nonuniform state of superconductors, Sov. Phys. JETP **20**, 762 (1965).

[6] H. A. Radovan, N. A. Fortune, T. P. Murphy, S. T. Hannahs, E. C. Palm, S. W. Tozer, D. Hall, Magnetic enhancement of superconductivity from electron spin domains, Nature **425**, 51 (2003).

[7] A. Bianchi, R. Movshovich, C. Capan, P. G. Pagliuso, J. L. Sarrao, Possible Fulde-Ferrell-Larkin-Ovchinnikov Superconducting State in $CeCoIn_5$, Phys. Rev. Lett. **91**, 187004 (2003).

[8] M. Kenzelmann, S. Gerber, N. Egetenmeyer, J. L. Gavilano, T. Strässle, A. D. Bianchi, E. Ressouche, R. Movshovich, E. D. Bauer, J. L. Sarrao, *et al.*, Evidence for a Magnetically Driven Superconducting Q Phase of $CeCoIn_5$, Phys. Rev. Lett **104**, 127001 (2010).

[9] J. Singleton, J. A. Symington, M. S. Nam, A. Ardavan, M. Kurmoo, P. Day, Observation of the Fulde-Ferrell-Larkin-Ovchinnikov state in the quasi-two-dimensional organic superconductor κ-(BEDT-TTF)$_2$Cu(NCS)$_2$, J. Phys. Condens. Matter **12**, L641 (2000).

[10] R. Lortz, Y. Wang, A. Demuer, P. H. M. Böttger, B. Bergk, G. Zwicknagl, Y. Nakazawa, J. Wosnitza, Calorimetric Evidence for a Fulde-Ferrell- Larkin-Ovchinnikov Superconducting State in the Layered Organic Superconductor κ-(BEDT-TTF)$_2$Cu(NCS)$_2$, Phys. Rev. Lett. **99**, 187002 (2007).

[11] B. Bergk, A. Demuer, I. Sheikin, Y. Wang, J. Wosnitza, Y. Nakazawa, R. Lortz, Magnetic torque evidence for the Fulde-Ferrell-Larkin-Ovchinnikov state in the layered organic superconductor κ-(BEDT-TTF)$_2$Cu(NCS)$_2$, Phys. Rev. B **83**, 064506 (2011).

[12] R. Beyer, B. Bergk, S. Yasin, J. A. Schlueter, J. Wosnitza, Angle-Dependent Evolution of the Fulde- Ferrell-Larkin-Ovchinnikov State in an Organic Superconductor, Phys. Rev. Lett. **109**, 027003 (2012).

[13] S. Tsuchiya, J.-i. Yamada, K. Sugii, D. Graf, J. S. Brooks, T. Terashima, S. Uji, Phase Boundary in

a Superconducting State of κ-(BEDT-TTF)$_2$Cu(NCS)$_2$: Evidence of the Fulde–Ferrell–Larkin–Ovchinnikov Phase, J. Phys. Soc. Jpn. **84**, 034703 (2015).
[14] C. C. Agosta, J. Jin, W. A. Coniglio, B. E. Smith, K. Cho, I. Stroe, C. Martin, S.W. Tozer, T. P. Murphy, E. C. Palm *et al.*, Experimental and semiempirical method to determine the Pauli-limiting field in quasi-two-dimensional superconductors as applied to κ-(BEDT-TTF)$_2$Cu(NCS)$_2$: Strong evidence of a FFLO state, Phys. Rev. B **85**, 214514 (2012).
[15] J. A. Wright, E. Green, P. Kuhns, A. Reyes, J. Brooks, J. Schlueter, R. Kato, H. Yamamoto, M. Kobayashi, S. E. Brown, Zeeman-Driven Phase Transition within the Superconducting State of κ−(BEDT−TTF)$_2$Cu(NCS)$_2$, Phys. Rev. Lett. **107**, 087002 (2011).
[16] H. Mayaffre, S. Kramer, M. Horvatić, C. Berthier, K. Miyagawa, K. Kanoda, V. F. Mitrović, Evidence of Andreev bound states as a hallmark of the FFLO phase in κ-(BEDT-TTF)$_2$Cu(NCS)$_2$, Nat. Phys. **10**, 928 (2014).
[17] K. Maki, T. Tsuneto, Pauli Paramagnetism and Superconducting State, Prog. Theor. Phys. **31**, 945 (1964).
[18] L. W. Gruenberg, L. Gunther, Fulde-Ferrell Effect in Type-II Superconductors, Phys. Rev. Lett. **16**, 996 (1966).
[19] L. G. Aslamazov, Influence of Impurities on the Existence of an Inhomogeneous State in a Ferromagnetic Superconductor, Sov. Phys. JETP **28**, 773 (1969).
[20] S. Takada, Superconductivity in a Molecular Field. II: Stability of Fulde-Ferrel Phase, Prog. Theor. Phys. **43**, 27 (1970).
[21] H. Shimahara, Fulde-Ferrell state in quasi-two-dimensional superconductors, Phys. Rev. B **50**, 12760 (1994).
[22] P. Burger, F. Hardy, D. Aoki, A. E. Böhmer, R. Eder, R. Heid, T. Wolf, P. Schweiss, R. Fromknecht, M. J. Jackson, *et al.*, Strong Pauli-limiting behavior of $H_{c2}$ and uniaxial pressure dependencies in $KFe_2As_2$, Phys. Rev. B **88**, 014517 (2013).
[23] D. A. Zocco, K. Grube, F. Eilers, T. Wolf, H. v. Löhneysen, Pauli-Limited Multiband Superconductivity in $KFe_2As_2$, Phys. Rev. Lett. **111**, 057007 (2013).
[24] J.-P. Reid, A. Juneau-Fecteau, R. T. Gordon, S. R. de Cotret, N. Doiron-Leyraud, X. G. Luo, H. Shakeripour, J. Chang, M. A. Tanatar, H. Kim, *et al.*, From d-wave to s-wave pairing in the iron-pnictide superconductor (Ba,K)$Fe_2As_2$, Supercond. Sci. Technol. **25**, 084013 (2012).
[25] P. Sullivan, G. Seidel, Steady-state, ac-temperature calorimetry, Phys. Rev. **173**, 679 (1968).
[26] R. Lortz, N. Musolino, Y. Wang, A. Junod, N. Toyota, Origin of the magnetization peak effect in the superconductor $Nb_3Sn$, Phys. Rev. B **75**, 094503 (2007).
[27] C. C. Agosta, N. A. Fortune, S. T. Hannahs, S. Gu, L. Liang, J.-H. Park, J. A. Schleuter, Calorimetric Measurements of Magnetic-Field-Induced Inhomogeneous Superconductivity Above the Paramagnetic Limit, Phys. Rev. Lett. **118**, 267001 (2017).
[28] Y. Wang, R. Lortz, Y. Paderno, V. Filippov, S. Abe, U. Tutsch, A. Junod, Specific heat and magnetization of a $ZrB_{12}$ single crystal: characterization of a type II/1 superconductor, Phys. Rev. B **72**, 024548 (2005).
[29] A. Gurevich, Upper critical field and the Fulde-Ferrel- Larkin-Ovchinnikov transition in multiband superconductors, Phys. Rev. B **82**, 184504 (2010).
[30] B. Jin, Fulde–Ferrell–Larkin–Ovchinnikov states in quasi-two-dimensional d + s-wave superconductors: Enhancement of the upper critical field, Physica C **468**, 2378 (2008).
[31] M. Takahashi, T. Mizushima, K. Machida, Multiband effects on Fulde-Ferrell-Larkin-Ovchinnikov

states of Pauli-limited superconductors, Phys. Rev. B **8**9, 064505 (2014).

[32] A. Ptok, D. Crivelli, The Fulde–Ferrell–Larkin–Ovchinnikov State in Pnictides, J. Low Temp. Phys. **172**, 226 (2013); A. Ptok, Influence of s± symmetry on unconventional superconductivity in pnictides above the Pauli limit – two-band model study, Eur. Phys. J. B 87, 2 (2014).

[33] T. Terashima, N. Kurita, M. Kimata, M. Tomita, S. Tsuchiya, M. Imai, A. Sato, K. Kihou, C.-H. Lee, H. Kito *et al.*, Fermi surface in $KFe_2As_2$ determined via de Haas–van Alphen oscillation measurements, Phys. Rev. B **87**, 224512 (2013).

[34] A. Ptok, Multiple phase transitions in Pauli-limited iron-based superconductors, J. Phys.: Condens. Matter **27**, 482001 (2015).

[35] F. Hardy, R. Eder, M. Jackson, D. Aoki, C. Paulsen, Th. Wolf, P. Burger, A. Böhmer, P. Schweiss, P. Adelmann *et al.*, Multiband Superconductivity in $KFe_2As_2$: Evidence for One Isotropic and Several Lilliputian Energy Gaps, J. Phys. Soc. Jpn. **83**, 014711 (2014).

[36] F. Hardy, A. E. Böhmer, D. Aoki, P. Burger, T. Wolf, P. Schweiss, R. Heid, P. Adelmann, Y. X. Yao, G. Kotliar *et al.*, Evidence of Strong Correlations and Coherence-Incoherence Crossover in the Iron Pnictide Superconductor $KFe_2As_2$. Phys. Rev. Lett. **111**, 027002 (2013).

[37] D. Saint-James, E. J. Thomas, G.Sarma, Type II Superconductivity (Pergamon, New York, 1969).

[38] T. Terashima, K. Kihou, M. Tomita, S. Tsuchiya, N. Kikugawa, S. Ishida, C.-H. Lee, A. Iyo, H. Eisaki, S. Uji, Hysteretic superconducting resistive transition in $Ba_{0.07}K_{0.93}Fe_2As_2$, Phys. Rev. B **87**, 184513 (2013).

[39] Field-induced superconducting phase of FeSe in the BCS-BEC cross-over, S. Kasahara, T. Watashige, T. Hanaguri, Y. Kohsaka, T. Yamashita, Y. Shimoyama, Y. Mizukami, R. Endo, H. Ikeda, K. Aoyama *et al.*, PNAS **111**, 16309 (2014).

**Supplementary information**

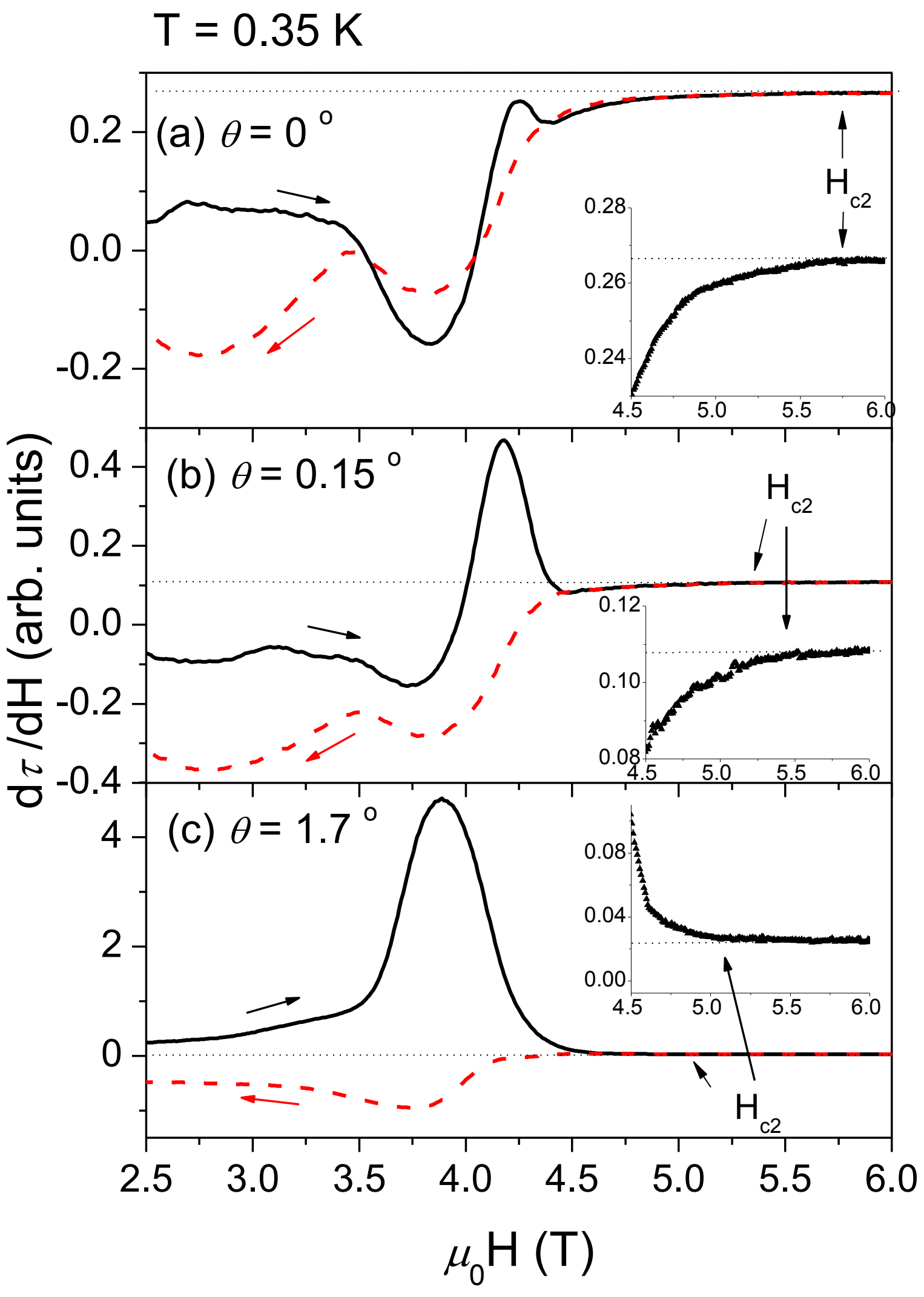

**Supplementary Figure 1.** The first-order derivative of the magnetic torque with respect to the applied field at (a) $\theta = 0°$, (b) $\theta = 0.15°$, and (c) $\theta = 1.7°$. Solid (dashed) lines mark data recorded with increasing (decreasing) field. The insets show enlarged regions at $H_{c2}$ (as marked by the vertical arrows). The dotted lines illustrate the constant normal state contribution.